\documentclass[aps,prl,preprintnumbers,amsmath,amssymb,twocolumn,superscriptaddress,showpacs,floatfix]{revtex4}
\usepackage{bbm}
\usepackage{color}
\usepackage{mathrsfs}
\usepackage{amsmath,amssymb,bm,comment}
\usepackage{hyperref}
\usepackage{graphicx}
\usepackage{epstopdf}
\usepackage{dcolumn}
\usepackage{bm}
\usepackage{lipsum}

\begin{document}

\title{Generalized Chiral Kinetic Equations}

\author{Shu-Xiang Ma}
\affiliation{Shandong Provincial Key Laboratory of Optical Astronomy and Solar-Terrestrial Environment,
School of Space Science and Physics, Shandong University, Weihai, Shandong 264209, China}

\author{Jian-Hua Gao}
\affiliation{Shandong Provincial Key Laboratory of Optical Astronomy and Solar-Terrestrial Environment,
School of Space Science and Physics, Shandong University, Weihai, Shandong 264209, China}

\begin{abstract}
We derive the generalized chiral kinetic equations which are applicable to  the fermions with arbitrary mass. We show how the dynamical magnetic-moment distribution function could lead to   spin polarization  and electric charge separation. We also show how the electric/magnetic moment distribution and pseudoscalar  distribution could be induced by  vorticity and electromagnetic field in global equilibrium.
\end{abstract}

\maketitle

{\it  Introduction.} ---
Recently some  novel spin phenomena have been observed in heavy-ion collisions \cite{Niida:2018hfw,STAR:2019erd,ALICE:2019aid}, such as opposite azimuthal angle dependence of hyperon polarization compared with  hydrodynamic simulation\cite{Karpenko:2016jyx,Becattini:2017gcx},
the inconsistency  between  vector meson and  hyperon \cite{Liang:2004ph,Liang:2004xn},
 and the differences of spin alignment between $K^*$  and $\phi$ mesons. These spin puzzles  have inspired lots of theoretical studies on the quantum kinetic theory  \cite{Gao:2019znl,Weickgenannt:2019dks,Hattori:2019ahi,Wang:2019moi,Li:2019qkf,Sheng:2020oqs,Guo:2020zpa,Sheng:2022ssd} which is supposed to be a promising approach to   tackle these problems associated with spin degree of freedom.

The chiral kinetic equation (CKE)  has been  established   to describe various chiral effects or spin effects for massless fermions ~\cite{Stephanov:2012ki,Son:2012zy,Chen:2012ca,Manuel:2013zaa,Manuel:2014dza,Chen:2014cla,Chen:2015gta,Hidaka:2016yjf,
Mueller:2017lzw,Huang:2018wdl,Gao:2018wmr,Liu:2018xip}.  However, the existing quantum kinetic theory for massive fermions takes very different form from the CKE  because the spin vector now  becomes dynamical quantity and  kinetic equations are  much more complicated than those in the chiral limit. Although there were a few  works \cite{Hattori:2019ahi,Wang:2019moi,Sheng:2020oqs,Guo:2020zpa} on demonstrating the smooth connection from the massive fermions to massless fermions, the final unification of  quantum kinetic equations for  massless and massive fermions is still expected. The unified kinetic equations would be essential to study  a system where a particle' mass emerges from zero smoothly, such as chiral symmetry breaking in heavy-ion collisions or the electroweak phase transion in the early universe.

In this Letter, we give  the generilized chiral kinetic equations (GCKE) which can be smoothly reduced to the CKE at chiral limit in a trivial way. After achieving these equations, we further use them to discuss the  spin polarization, charge separation, electric/magnetic moment distribution and pseudoscalar condensation which are very relevant to the phenomena in  heavy-ion collisions.

{\it  Wigner function formalism.} ---
We will derive the GCKE by using the covariant Wigner function formalism ~\cite{Heinz:1983nx,Elze:1986qd,Vasak:1987um}.
The Wigner function $W(x,p)$ is defined as
the ensemble average of the bilinear Dirac fields:
\begin{eqnarray*}
\label{wigner}
W_{\alpha\beta}
=\int\frac{d^4 y}{(2\pi)^4}
e^{-ip\cdot y}\langle \bar\psi_\beta(x_+)U(x_+,x_-)
\psi_\alpha(x_-)\rangle, \
\end{eqnarray*}
where $U(x_+,x_-)  =e^{-i \int_{x_-}^{x_+} dz^\mu A_\mu (z)}$
is the gauge link along the  straight line connecting the Dirac field
with spinor index $\alpha$ at point $x_-=x- y/2$ to the conjugate field with spinor index $\beta$ at point  $x_+=x+y/2$.
We have neglected the path-order operator in the gauge link because  the background gauge field approximation
would be enough for us to exhibit the procedure to arrive at the GCKE. For simplicity, we have absorbed the electric charge $e$ into the gauge potential $A_\mu$. The Wigner function is a $ 4 \times 4$ matrix in spinor space and can be  decompose into  scalar
 $\mathscr{F}$, pseudoscalar $\mathscr{P}$, vector $\mathscr{V}_{\mu}$, axial vector $\mathscr{A}_{\mu}$ and antisymmetric tensor $\mathscr{S}_{\mu\nu}$ components:
\begin{align*}
\label{decomposition}
W&=\frac{1}{4}[\mathscr{F}+i\gamma^5 \mathscr{P}+\gamma^\mu \mathscr{V}_\mu +\gamma^5\gamma^\mu \mathscr{A}_\mu
+\frac{1}{2}\sigma^{\mu\nu} \mathscr{S}_{\mu\nu}].
\end{align*}
Since we want to  put the final equations in a similar form to the CKE, we will use the chirality basis
$\mathscr{J}_{s}^\mu = (\mathscr{V}^{\mu} + s\mathscr{A}^{\mu} )/2$
where $s= + 1$ and $s= - 1$ denotes the  right- and left-hand chirality, respectively.  The Dirac equations can result in
two groups of  coupled equations:
\begin{eqnarray}
\label{Js-cs}
p_\mu\mathscr{J}^\mu_s &=& \frac{1}{2} m  \mathscr{F},\\
\label{Js-ev1}
 \nabla_\mu\mathscr{J}^\mu_s &=& -   m s\mathscr{P},\\
\label{Js-ev2}
2s(p^\mu \mathscr{J}_{s}^\nu - p^\nu \mathscr{J}_{s}^\mu )
 &=& - \epsilon^{\mu\nu\rho\sigma}\nabla_\rho \mathscr{J}_{s\sigma} + m \tilde{\mathscr{S}}^{\mu\nu}
\end{eqnarray}
together with
\begin{eqnarray}
\label{F-cs}
p^\mu\mathscr{F}+\frac{1}{2} \nabla_\nu\mathscr{S}^{\mu\nu} &=& m \sum_{s}\mathscr{J}^\mu_{s},\\
\label{P-cs}
p^\mu\mathscr{P}+\frac{1}{2}\nabla_\nu {\tilde{\mathscr{S}}}^{\mu\nu} &=&0,\\
\label{P-ev}
\nabla_\mu\mathscr{P}-2p^\nu\tilde{\mathscr{S}}_{\mu\nu}&=& 2m \sum_{s} s \mathscr{J}^\mu_{s},\\
\label{F-ev}
\nabla_\mu\mathscr{F}-2p^\nu\mathscr{S}_{\mu\nu}&=&0.
\end{eqnarray}
These equations are valid up to the first order of $\hbar$ and the operator $\nabla^\mu=\partial^\mu_x - F^{\mu\nu}\partial_\nu^p$ and
 ${\tilde{\mathscr{S}}}^{\mu\nu} \equiv \epsilon^{\mu\nu\rho\sigma}\mathscr{S}_{\rho\sigma}/2$ with
 the convention $\epsilon^{0123}=1$.

{\it  Disentangling the Wigner equations.} ---
 We note  that in the Eqs. (\ref{Js-cs})-(\ref{F-ev}) not all the Wigner functions are independent  and some Wigner functions can be expressed in terms of other Wigner fucntions. For example, the first group  equations can express all  $\mathscr{F}$, $\mathscr{P}$ and ${\mathscr{S}}^{\mu\nu}$ in terms of $\mathscr{J}^\mu_s$. However in these expressions the mass term in the denominator would be unavoidable and this makes the chiral limit very subtle.
In order to make the transition from massive to massless smoothly in a trivial way, we will avoid putting mass in  the denominator.
Let us introduce a constant timelike 4-vector $n^\mu$ with $n^2=1$. Then we can decompose any vector $X^\mu$ into the timelike component $X_n =n\cdot X$ which is parallel to $n^\mu$ and the spacelike component $\bar X^\mu = \Delta^{\mu\nu}X_\nu$ which perpendicular to $n^\mu$ with
$\Delta_{\mu\nu}=g_{\mu\nu}-n_\mu n_\nu$. For example, $p^\mu = p_n n^\mu + \bar p^\mu $, $\nabla^\mu =n^\mu \nabla_n  + \bar\nabla^\mu$, and
 $\mathscr{J}_{s}^{\mu} =\mathscr{J}_{sn} n^\mu +\bar{\mathscr{J}}_{s}^\mu$. Similarly the antisymmetric  tensor can be
 also decomposed as
\begin{eqnarray}
\label{Fmunu}
F^{\mu\nu}&=&E^\mu n^\nu -E^\nu n^\mu -\bar\epsilon^{\mu\nu\sigma} B_\sigma,\\
\label{Smunu}
\mathscr{S}^{\mu\nu}&=& \mathscr{K}^\mu n^\nu - \mathscr{K}^\nu n^\mu - \bar\epsilon^{\mu\nu\sigma} \mathscr{M}_{\sigma},
\end{eqnarray}
where the tensor $\bar\epsilon^{\mu\nu\sigma}\equiv \epsilon^{\mu\rho\nu\sigma}n_\rho$ has been defined for convenience,  $E^\mu$/$B^\mu$ denotes the electric/magnetic field in a frame with velocity $n^\mu$ and$\mathscr{K}^\mu$/$\mathscr{M}^\mu$ denotes the electric/magnetic moment distribution function in the same frame. Now we can always put $p_n$ in the denominator which have smooth limit when the mass of the fermions approaches to zero. From the  timelike-spacelike $n\bar \nu$-components of $\mu\nu$ in Eq.(\ref{Js-ev2}), we can express $\bar{ \mathscr{J}}_{s}^{\mu}$ in terms of $\mathscr{J}_{sn}$ and $\mathscr{M}^{\mu}$:
\begin{eqnarray}
\label{barJs-Jsn-M-1}
\bar{ \mathscr{J}}_{s}^{\mu}
 =(\Delta^{\mu\sigma}  +\frac{s }{2p_n}\bar\epsilon^{\mu\rho\sigma}\nabla_\rho )
 (\frac{\bar  p_\sigma}{p_n} \mathscr{J}_{sn} -\frac{s  m }{2p_n} \mathscr{M}_{\sigma} )
\end{eqnarray}
Similarly, from timelike $n$-component of Eq.(\ref{F-cs}), we
can express $\mathscr{F}$ as
\begin{eqnarray}
\label{F-Jsn-1}
\mathscr{F} = \frac{m}{p_n} \sum_{s}\mathscr{J}_{sn}
+\frac{1}{2p_n}\bar \epsilon^{\mu\alpha\beta} \nabla_\mu( \frac{p_\alpha}{p_n}    \mathscr{M}_{\beta})
\end{eqnarray}
and from the timelike $n$-component of Eq.(\ref{P-cs}) we have
\begin{eqnarray}
\label{P-Jsn-1}
\mathscr{P} = \frac{1}{2p_n}\nabla_\mu {\mathscr{M}}^{\mu}
\end{eqnarray}
The Wiger function $\mathscr{K}_{\mu}$ can also be expressed in terms of $\mathscr{J}_{sn}$ and $\mathscr{M}^{\mu}$ by using the $n$-component of   Eq.(\ref{F-ev}) together with (\ref{F-Jsn-1}):
\begin{eqnarray}
\label{K-M-1}
 \mathscr{K}_{\mu} =  \frac{1}{p_n} \bar \epsilon_{\mu\alpha\beta} p^\alpha  \mathscr{M}^{\beta}
 +\frac{1}{2p_n}\bar\nabla_\mu (\frac{m}{p_n} \sum_{s}\mathscr{J}_{sn})
\end{eqnarray}
Then  antisymmetric tensor  Wigner function follows as
\begin{eqnarray}
\label{S-M}
\mathscr{S}^{\mu\nu}
&=&\frac{1}{p_n} \epsilon^{\mu\nu\alpha\beta}p_\alpha \mathscr{M}_{\beta}\nonumber\\
& &+\frac{m}{2p_n}(n_\nu \bar\nabla_\mu - n_\mu \bar\nabla_\nu)( \frac{1}{p_n} \sum_{s}\mathscr{J}_{sn})
\end{eqnarray}
The above expressions have explicitly shown  that we can choose $\mathscr{J}_{sn}$ and $\mathscr{M}^{\mu}$ as the independent distribution functions and all the other Wigner functions can be derivative from them. From the definition (\ref{Smunu}), we know that the magnetic moment distribution $\mathscr{M}^{\mu}$ must be  pure space-like, i.e., $n_\mu \mathscr{M}^{\mu}=0$. Hence only
three components of $\mathscr{M}^{\mu}$ are independent. It is convenient to decompose $\mathscr{M}^{\mu}$ into
$\mathscr{M}^{\mu}= \mathscr{M}^{\mu}_\parallel + \mathscr{M}^{\mu}_\perp$
where $\mathscr{M}^{\mu}_\parallel $ denotes the longitudinal component  parallel to the momentum $\bar p^\mu$ and $\mathscr{M}^{\mu}_\perp $ denotes the transverse component orthogonal to the  momentum $\bar p^\mu$.  If we introduce the transverse projector operator $\Delta^{\mu\nu}_\perp = \Delta^{\mu\nu} - \bar p^\mu \bar p^\nu /\bar p^2$, then the transverse component of a four-vector $X^\mu$ can be obtained directly by
 $X^\mu_\perp = \Delta^{\mu\nu}_\perp X_\nu$.
It is easy to verify that the timelike component of Eq.(\ref{P-ev}) actually further constrain $\mathscr{M}^{\mu}_\parallel$ as
\begin{eqnarray}
\label{M-p-1}
\mathscr{M}^{\mu}_\parallel = \frac{m \bar p^\mu }{\bar p^2} \sum_{s} s \mathscr{J}_{sn}
\end{eqnarray}
which means that the longitudinal magnetic moment distribution is determined by the imbalance of
the chirality distribution and will vanish as the mass approaches zero.
Hence only  $\mathscr{M}^{\mu}_\perp$  are independent. Now the transport equation for $\mathscr{J}_{sn}$ is obtained by substituting
expressions (\ref{barJs-Jsn-M-1}) and (\ref{P-Jsn-1}) into Eq.(\ref{Js-ev1}) and the transport equation  for $\mathscr{M}^{\mu}$ is obtained
by substituting expressions (\ref{K-M-1}) and (\ref{P-Jsn-1}) into Eq.(\ref{P-cs}). These equations are totally entangled with each other:
\begin{widetext}
\begin{eqnarray}
\label{Jsn-kinetic-eq-1-a1}
p^\mu \nabla_\mu ( \frac{\mathscr{J}_{sn}}{p_n})
&=&-\frac{s}{2}\bar\epsilon^{\mu\rho\sigma}\nabla_\mu [\frac{1}{p_n}\nabla_\rho
(\bar  p_\sigma\frac{\mathscr{J}_{sn}}{p_n}  -\frac{s  m }{2}\cdot\frac{ \mathscr{M}_{\sigma}}{p_n} ) ]
+\frac{ m s}{2p_n}E_\mu \frac{{\mathscr{M}}^{\mu}}{p_n},\\
\label{M-kinetic-eq-1-a1}
p^\nu \nabla_\nu (\frac{\mathscr{M}^{\mu}}{p_n})
&=&\frac{1}{p_n}[ \bar p^\mu E^\nu - p_n \bar\epsilon^{\mu\nu\alpha}B_\alpha ]\frac{\mathscr{M}_\nu}{p_n}
- \frac{m}{2} \bar\epsilon^{\mu\nu\rho}\nabla_\nu [\frac{1}{p_n}\nabla_\rho
( \sum_{s}\frac{\mathscr{J}_{sn}}{p_n}) ].
\end{eqnarray}
\end{widetext}
In additon,  $\mathscr{J}_{sn}$ and $\mathscr{M}^{\mu}$ will be further constrained by other remained Wigner equations.
Inserting (\ref{barJs-Jsn-M-1}) and (\ref{F-Jsn-1}) into (\ref{Js-cs}) leads to
\begin{eqnarray}
\label{Jsn-onshell}
(p^2-m^2)\frac{\mathscr{J}_{sn}}{p_n}
=\frac{s}{p_n}B^\mu (\bar  p_\mu \frac{\mathscr{J}_{sn}}{p_n}  -\frac{s  m }{2}\cdot\frac{ \mathscr{M}_{\mu}}{p_n} )
\end{eqnarray}
which is just the modified on-shell condition for $\mathscr{J}_{sn}$. The similar on-shell condition for $\mathscr{M}^{\mu}$
can be obtained by inserting (\ref{K-M-1}) and (\ref{F-Jsn-1}) into (\ref{F-ev})
\begin{eqnarray}
\label{Mmu-onshell}
(p^2-m^2)\frac{{\mathscr{M}}_{\mu}}{p_n}
= \frac{m}{p_n}B_\mu \sum_{s} \frac{\mathscr{J}_{sn}}{p_n}
\end{eqnarray}
Now we have succeeded in disentangling the original Wigner functions which satisfy Wigner  equations (\ref{Js-cs}-\ref{F-ev}) into the independent functions $\mathscr{J}_{sn}$ and $\mathscr{M}^{\mu}$ which satisfy equations
(\ref{Jsn-kinetic-eq-1-a1},\ref{M-kinetic-eq-1-a1})
together with the constraint conditions (\ref{M-p-1}), (\ref{Jsn-onshell}) and (\ref{Mmu-onshell}).
All the other Wigner functions can be determined from  $\mathscr{J}_{sn}$ and $\mathscr{M}^{\mu}$ by Eqs.(\ref{F-Jsn-1}-\ref{K-M-1}). We can
 verify that  all the other equations which we did not used yet such as spacelike-spacelike components of Eq.(\ref{Js-ev2}) and  the spacelike components of Eq.(\ref{F-cs}) or (\ref{P-cs}) are all fulfilled automatically.

{\it Generalized chiral kinetic equations.} --- Now we should eliminate the final constraints (\ref{Jsn-onshell}) and  (\ref{Mmu-onshell})
and obtain the final kinetic equations without any extra constraints. As we mentioned above that these equations imply the
generalized mass-shell conditions. The constraint condition \ref{Jsn-onshell}
indicates  that  $\mathscr{J}_{sn}$ must take the form
\begin{eqnarray}
\label{calJsn}
\mathscr{J}_{sn} &=& p_n \mathcal{J}_{sn}\delta(p^2-m^2)\nonumber\\
& &-s  B^\mu (  \bar p_\mu \mathcal{J}_{sn}
- \frac{s m}{2}\mathcal{M}_{\mu})\delta'(p^2-m^2)\hspace{1cm}
\end{eqnarray}
and the constraint condition \ref{Mmu-onshell} leads to
\begin{eqnarray}
\label{calM}
{\mathscr{M}}_{\mu} &=& p_n {\mathcal{M}}_{\mu}\delta(p^2-m^2)\nonumber\\
& &-  m  B_\mu \sum_{s}\mathcal{J}_{sn}\delta'(p^2-m^2)
\end{eqnarray}
where we have introduced new distribution functions $ \mathcal{J}_{sn}$ and ${\mathcal{M}}_{\mu}$ to replace the original functions
$\mathscr{J}_{sn}$ and $\mathscr{M}^{\mu}$.
As  we decompose $ {\mathscr{M}}_{\mu}$ into longitudinal and transverse parts above and
only transverse parts are independent, we can also write  ${\mathcal{M}}_{\mu}$ in terms of longitudinal and transverse parts,
$\mathcal{M}^{\mu}= \mathcal{M}^{\mu}_\parallel + \mathcal{M}^{\mu}_\perp$
where the longitudinal part is gives by
\begin{eqnarray}
\label{calM-p-1}
\mathcal{M}^{\mu}_\parallel = \frac{m \bar p^\mu }{\bar p^2} \sum_{s} s \mathcal{J}_{sn}
\end{eqnarray}

Substituting (\ref{calJsn}) and (\ref{calM}) into (\ref{Jsn-kinetic-eq-1-a1}) and (\ref{M-kinetic-eq-1-a1}) gives rise to
the transport equation for  $ \mathcal{J}_{sn}$ and ${\mathcal{M}}_{\mu}$. However
these equations involve the singular $\delta$ function and are not suitable for the numerical simulation. After  integrating over the timelike $n$-component of the momentum, we obtain the final GCKE  for the positive particles with momentum $\bar p$ and chirality $s$:
\begin{widetext}
\begin{eqnarray}
\label{GCKE-Jsn}
& &(1-{s{B\cdot \Omega}}) \partial^x_n \tilde{\mathcal{J}}_{sn}
+[(1-2{s{B\cdot \Omega}})\bar v^{\mu}
+ s \bar\epsilon^{\mu\rho\sigma} \Omega_\rho E_\sigma ] \bar\partial^x_\mu \tilde{\mathcal{J}}_{sn} \nonumber\\
& &+[ E^\mu + \mathcal{E}_{\bar p} \bar\partial_x^\mu (B\cdot \Omega)
-(1 - 2{s{B\cdot \Omega}})
\bar\epsilon^{\rho\sigma\mu}\bar v_{\rho} B_\sigma
 -  s ( E\cdot B)\Omega^\mu ]
 \partial^p_\mu  \tilde{\mathcal{J}}_{sn}\nonumber\\
&=&  \frac{ m s}{2\mathcal{E}_{\bar p}^2} (1 - {s{B\cdot \Omega}}) E_\mu \tilde{\mathcal{M}}^{\mu}
-\frac{ m }{4\mathcal{E}_{\bar p}^3}(\bar v^\mu B^\nu + \bar\epsilon^{\mu\rho\nu} E_\rho)
\bar\partial^x_\mu \tilde{\mathcal{M}}_\nu \nonumber\\
& &- \frac{m}{4\mathcal{E}_{\bar p}^3} [( E\cdot B)\Delta^{\mu\nu}
- B^\mu E^\nu - B^\nu \bar\epsilon^{\mu\rho\sigma} \bar v_\rho B_\sigma
-\mathcal{E}_{\bar p} \bar\partial_x^\mu B^\nu]\bar\partial^p_\mu \tilde{\mathcal{M}}_{\nu},\\
\label{GCKE-Mmu}
& &\partial^x_n \tilde{\mathcal{M}}^{\mu}_{\perp}
+\bar v^{\nu} \bar\partial^x_\nu \tilde{\mathcal{M}}^{\mu}_{\perp}
+ (E^\nu -  \bar\epsilon^{\rho\sigma\nu}\bar v_{\rho} B_\sigma )
\partial_\nu^p \tilde{\mathcal{M}}^{\mu}_{\perp}
+ \frac{\mathcal{E}_{\bar p} }{\bar p^2}(\bar v^\mu E^\nu + \bar v^2 \bar\epsilon^{\mu\nu\alpha}B_\alpha)
\tilde{\mathcal{M}}_{\perp\nu}\nonumber\\
&=&-\frac{ m  }{ \bar p^2}E^\mu_\perp \sum_{s} s \tilde{\mathcal{J}}_{sn}
-\frac{m}{2\mathcal{E}_{\bar p} \bar p^2}
[(\bar v^\mu \bar\epsilon^{\sigma\nu\rho}-\bar v^\sigma \bar\epsilon^{\mu\nu\rho})
\bar v_\sigma E_\rho -  \bar v^2  B_\perp^\mu \bar v^\nu]\partial^x_\nu  \sum_s \tilde{\mathcal{J}}_{sn}\nonumber\\
& &-\frac{m}{2\mathcal{E}_{\bar p}^3 } [B^\mu_\perp E^\nu  - (E\cdot B) \Delta^{\mu\nu}_\perp
 +   B_\perp^\mu  \bar\epsilon^{\rho\sigma\nu}  \bar v_\rho B_\sigma
 + \mathcal{E}_{\bar p} (\partial_x^\nu B^\mu_\perp)]
\bar \partial_\nu^p  \sum_s \tilde{\mathcal{J}}_{sn}.
\end{eqnarray}
\end{widetext}
where $\mathcal{E}_{\bar p}=\sqrt{m^2-\bar p^2}$ denotes the free Dirac particle's energy,
$\bar v^\mu = \bar p^\mu /\mathcal{E}_{\bar p}$ denotes the free Dirac particles's spatial velocity,
$\Omega^\mu = \bar p^\mu/2\mathcal{E}_{\bar p}^3$  is the Berry curvature for massive Dirac fermion,
and $(1-{s{B\cdot \Omega}})$ is the modification to the invariant phase space for massive Dirac fermion just as the case of massless fermions.
The GCKE  for the negative particles with momentum $\bar p$ and chirality $-s$ can be easily obtained by replacing $e$ implicit in $E^\mu$ or $B^\mu$
by $-e$ directly.
Note that we have redefined new distribution functions $\tilde{\mathcal{J}}_{sn}$ and $\tilde{\mathcal{M}}^{\mu}_{\perp}$ which are
related to ${\mathcal{J}}_{sn}$ and ${\mathcal{M}}^{\mu}_{\perp}$ by
\begin{align*}
\tilde{\mathcal{J}}_{sn} \equiv&\,(  \mathcal{J}_{sn}
+\frac{sB\cdot \bar p}{2 \mathcal{E}_{\bar p}^2}\mathcal{J}'_{sn}
-\frac{m B^\nu}{4\mathcal{E}_{\bar p}^2}\mathcal{M}'_\nu )_{p_n=\mathcal{E}_{\bar p}},\\
\tilde{\mathcal{M}}^{\mu}_{\perp} \equiv &\,( \mathcal{M}^{\mu}_{\perp}
+ \frac{ m B_\perp^\mu}{2\mathcal{E}_{\bar p}^2}\sum_s \mathcal{J}'_{sn})_{p_n=\mathcal{E}_{\bar p}},
\end{align*}
where  $\mathcal{J}'_{sn}\equiv\partial_{p_n}\mathcal{J}_{sn}$ and
 $\mathcal{M}'_\nu\equiv\partial_{p_n}\mathcal{M}_\nu$ and all the timelike component $p_n$ has been fixed by the onshell condition
$p_n=\mathcal{E}_{\bar p}$ after integrating over $\delta(p^2-m^2)$ or $\delta'(p^2-m^2)$.
In general, after integrating over $p_n$, the derivative terms are independent on $\mathcal{J}_{sn}$ and $\mathcal{M}_\nu$ at all.
Fortunately, they always appears as a whole in terms of $\tilde{\mathcal{J}}_{sn}$ and $\tilde{\mathcal{M}}^{\mu}_{\perp} $. Actually
all the Wigner functions after integrating over $p_n$ depend only on $\tilde{\mathcal{J}}_{sn}$ and $\tilde{\mathcal{M}}^{\mu}_{\perp} $. Correspondingly, the final physical quantities such as charge currents or energy-momentum tensor can only depend on $\tilde{\mathcal{J}}_{sn}$ and $\tilde{\mathcal{M}}^{\mu}_{\perp} $.

The GCKE (\ref{GCKE-Jsn}) and  (\ref{GCKE-Mmu})  are the main result of this work. It is obvious that the right hand of the two equations
will both vanish at chiral limit and the two equations are decoupled with each other.  The first equation will give the CKE
and the second equation gives the kinetic equation for $\tilde{\mathcal{M}}_{\perp\nu}$. We note that when we perform the chiral limit,
we will not encounter the mass in the denominator so that the chiral limit is trivial in our approach.

Now we can use our formalism to consider the   spin polarization effect.
We can analyze the spin  polarization  by  the space-like component  of axial Wigner function. After integrating over $p_n$, we obtain
\begin{widetext}
\begin{equation}
 \label{barJ-1-int}
\int dp_n {\bar{ \mathscr{A}}}^{\mu}
 =\frac{ 1}{2}\bar v^\mu\Delta\tilde{\mathcal{J}}_{n}
 + \frac{ 1 }{{\mathcal{E}}_{\bar p} }[
 \mathcal{E}_{\bar p}\bar\epsilon^{\mu\rho\sigma}\partial^x_\rho
- 2\bar v^\mu B^\sigma - \bar\epsilon^{\mu\rho\sigma}E_\rho
+\mathcal{E}_{\bar p}(B^\sigma \bar\partial_p^\mu - B^\mu \bar\partial_p^\sigma)]
\frac{ \bar p_\sigma \tilde{\mathcal{J}}_{n}}{4\mathcal{E}_{\bar p}^2}
 -\frac{  m ({\mathcal{E}}_p^3 \tilde{\mathcal{M}}^{\mu}
- m B^\mu \tilde{\mathcal{J}}_{n})}{2 {\mathcal{E}}_p^4 }
\end{equation}
\end{widetext}
where $\tilde{\mathcal{J}}_{n}= \sum_s \tilde{\mathcal{J}}_{sn}$ represents the  sum of
different chirality distributions  while   $\Delta\tilde{\mathcal{J}}_{n}= \sum_s s\tilde{\mathcal{J}}_{sn}$ denotes the imbalance between two different distributions. Similar unintegrated results has been discussed in Refs.\cite{Weickgenannt:2019dks} and \cite{Liu:2021nyg}.
The first term in square brackets includes the polarization effect from vorticity, acceleration, or shear tensor. The second term denotes the polarization due to magnetic field and the third term means the spin Hall effect \cite{Liu:2020dxg}.  The final term in square brackets is total derivative and will not contribution after integrating over the momentum. However, it might contribute to the momentum dependence of the spin polarization.  In addition to these polarization effects, we note that  the last term implies a possible generation mechanism due to the dynamical magnetic moment distribution.  It is obvious that this  contribution is proportional to the particle's mass $m$ and might be very relevant to the hyperon's polarization and spin alignment of vector mesons associated with the strange quark in  heavy-ion collisions. How this term contributes to the polarization  quantitatively deserve further investigation  in the future and might shed light on the spin puzzles  in heavy-ion collisions.

Similarly, we can study  the electric charge separation by the space-like component of the vector Wigner function. After integrating over $p_n$, we obtain
\begin{widetext}
\begin{equation}
 \label{barJ-1-int}
\int dp_n {\bar{ \mathscr{V}}}^{\mu}
 =\frac{1}{2}\bar v^\mu\tilde{\mathcal{J}}_{n}
+\frac{1}{\mathcal{E}_{\bar p}}[\mathcal{E}_{\bar p}\bar\epsilon^{\mu\rho\sigma}\partial^x_\rho
-2\bar v^\mu B^\sigma - {\bar\epsilon^{\mu\rho\sigma}E_\rho}
+\mathcal{E}_{\bar p}(B^\sigma \bar\partial_p^\mu - B^\mu \bar\partial_p^\sigma)]
\frac{ \bar p_\sigma \Delta\tilde{\mathcal{J}}_{n} -m\tilde{\mathcal{M}}_\sigma}{4\mathcal{E}_{\bar p}^2}
\end{equation}
\end{widetext}
In square brackets, the first and second terms
correspond to well-known chiral vortical effect and chiral magnetic effect, respectively.
The difference from the massless  case are quite clear. In addition to the mass effect in $\mathcal{E}_{\bar p}$, the major difference is due to the contribution from the dynamical magnetic moment distribution $\tilde{\mathcal{M}}_\sigma$.

{\it Global equilibrium solution.} ---
We will also apply our formalism to  determine various physical quantities in global equilibrium when
  the vorticity and electromagnetic field are present. In global equilibrium, the fluid velocity $u^\mu$ (with normalization $u^2=1$), chemical potential $\mu$ and temperature $T$ of the system must satisfy the constraints \cite{Yang:2022ksq}:
\begin{eqnarray*}
\label{C-EQ}
\partial_\mu \beta_\nu + \partial_\nu \beta_\mu =0  ,\ \ \
\partial_\mu \bar \mu = -F_{\mu\nu}\beta^\nu
\end{eqnarray*}
where $\beta^\mu = \beta u^\mu $ with $\beta = 1/T$. These constraints lead to the fact that the the thermal vorticity tensor
$\Omega^{\mu\nu}=(\partial^\mu \beta^\nu - \partial^\nu \beta^\mu)/2$ must be constant and the relation
  ${F_{\lambda}}^\mu \Omega^{\nu\lambda}-{F_{\lambda}}^\nu \Omega^{\mu\lambda}=0$. For simplicity, we also assume the electromagnetic field imposed on the fluid is constant. Just like the antisymmetric tensor $F^{\mu\nu}$ and $\mathscr{S}^{\mu\nu}$, we can decompose   the vorticity tensor  as
\begin{eqnarray}
\Omega^{\mu\nu} = \beta\left(\varepsilon^\mu n^\nu - \varepsilon^\nu n^\mu -\bar\epsilon^{\mu\nu\sigma}\omega_\sigma\right)
\end{eqnarray}
where we  refer to $\omega^\mu$  as vorticity vector and  $\varepsilon^\mu$ as acceleration vector.
 In general, we can not obtain  a unique specific solution from the kinetic equations. However if we require such solution must smoothly lead to the one at chiral limit which has been already known, then we can determine them uniquely. We will assume the the distribution functions are unpolarized and magnetic moment distribution vanish when the vorticity and electromagnetic field are absent. Then we can assume the Wigner function $\mathscr{J}_{sn} $  as
\begin{eqnarray}
\label{Jsn-onshell-1-b-EQ-2}
\mathscr{J}_{sn} &=& ( p_n f
 +  \frac{s}{2}\omega\cdot \bar p\,  f' )\delta(p^2-m^2)\nonumber\\
& & -s  B\cdot  \bar p f\delta'(p^2-m^2)
\end{eqnarray}
where  the unpolarized distribution function $f$ does not depend on the chirality or spin and is given by
\begin{equation}
\label{calJ-n}
f= \frac{1}{4\pi^3}
[\theta (p_0) n(y)  +\theta (-p_0)  n(-y)
-\theta (-p_0)]
\end{equation}
where  $n(y)=(1+ e^{y})^{-1}$ is Fermi-Dirac distribution and $y=\beta\cdot p-\bar\mu$ with $\bar \mu \equiv \mu /T$ being
 the scaled charge chemical potential by the temperature $T$. We note that
 the result (\ref{Jsn-onshell-1-b-EQ-2}) will recover the solution at chiral limit from CKE.
Substituting this expression into Eqs. (\ref{Jsn-kinetic-eq-1-a1}) and (\ref{M-kinetic-eq-1-a1}), we find ${\mathscr{M}}^{\mu}_{\perp} $ should be
\begin{eqnarray}
\label{M-onshell-1-b-EQ-2}
{\mathscr{M}}^{\mu}_{\perp} = m\omega_\perp^\mu f^{\prime} \delta(p^2-m^2)- 2 m B^\mu_\perp f\delta'(p^2-m^2),
\end{eqnarray}
Once these fundamental functions are known, the other Wigner functions can be derived directly from the preceding results.
For example, we can have
\begin{eqnarray}
{ \mathscr{J}}_{s}^{\mu}
&=&( p^\mu f -\frac{s}{2}  \tilde\Omega^{\mu\nu}p_\nu  f') \delta(p^2-m^2)\nonumber\\
& & + s \tilde F^{\mu\nu}p_\nu  f \delta'(p^2-m^2)
 \end{eqnarray}
which is consistent with the results in \cite{Fang:2016vpj} and \cite{Lin:2018aon}.  Here we will focus on the other Wigner functions.
The scalar condensation can be given from (\ref{F-Jsn-1}) directly,
\begin{eqnarray}
{\mathscr{F}}
&=& 2 m f\delta(p^2-m^2)
\end{eqnarray}
which implies that the vorticity or electromagnetic field will not induce scalar condensation for unpolarized systems.  From (\ref{S-M}), we can obtain the antisymmetric moment tensor,
\begin{eqnarray}
{\mathscr{S}}^{\mu\nu}
=m\Omega^{\mu\nu}f^{\prime}\delta(p^2-m^2) - 2 m  F^{\mu\nu}f\delta'(p^2-m^2).
\end{eqnarray}
Integrating over momentum gives rise to
\begin{eqnarray}
\label{S-O-F}
\int d^4 p {\mathscr{S}}^{\mu\nu}
=\kappa_\Omega  T \Omega^{\mu\nu}+\kappa_F F^{\mu\nu},
\end{eqnarray}
where the coefficients $\kappa_\Omega $ and $\kappa_F$ are given by
\begin{eqnarray*}
\kappa_\Omega &=&  - \frac{m }{2\pi^2}\int dp \left[ n(y_+) -n(y_-) \right],\\
\kappa_F &=& -  \frac{m}{2\pi^2}\int \frac{d p}{\mathcal{E}_{\bar p}}\left[ n(y_+) + n(y_-) \right] +  \kappa_F^{\textrm{vac}}
\end{eqnarray*}
where $y_\pm = \beta \mathcal{E}_{\bar p} \mp \bar \mu$ and $\kappa_F^{\textrm{vac}} $ denotes the contribution from the last vacuum
term in Eq.(\ref{calJ-n}),
\begin{equation*}
\kappa_F^{\textrm{vac}}
=\frac{m}{4\pi^2}( \frac{2}{\epsilon}-\ln m -\gamma +\mathcal{O}(\epsilon))
\end{equation*}
where we have  calculate the integral $\int d^4 p$ by using the dimensional regularization $\int d^{4-\epsilon} p$.
The results (\ref{S-O-F}) imply that the electric moment distribution can be induced by the acceleration  vector $\varepsilon^\mu$
and electric field $E^\mu$ and  the magnetic moment distribution can be induced by the vorticity  vector $\omega^\mu$ and magnetic field $B$.

As we mentioned at the beginning that  our equations are only valid up to  the first order of $\hbar $, at this order there is no  pseudoscalar condensation vanishes from Eq.(\ref {P-Jsn-1}). Fortunately,  the equation (\ref {P-Jsn-1})   actually hold even at the second order  and  lead to the second-order  pseudoscalar condensation  as the following
\begin{eqnarray*}
\label{P-Jsn-2-EQ-1}
\int d^4 p \mathscr{P}
=C_{\Omega}T^2\tilde \Omega^{\mu\nu}\Omega_{\mu\nu}
+C_{M}T \tilde \Omega^{\mu\nu}F_{\mu\nu}+C_{F} \tilde F^{\mu\nu}F_{\mu\nu}
\end{eqnarray*}
where $C_{\Omega} = - \frac{1}{8}\frac{\partial \kappa_\Omega}{\partial \mu},\ \ \
C_{M} = -\frac{1}{4}\frac{\partial \kappa_F}{\partial \mu },$ and
\begin{eqnarray*}
C_{F} &=& -\frac{m}{16\pi^2}\int
\frac{dp}{ p^2 \mathcal{E}_{\bar p}} \left[n(y_+) + n(y_-)  -1\right]
\end{eqnarray*}
These results are consistent with the ones obtained in Ref. \cite{Fang:2016uds} for  $C_{M} $ and $C_{F} $ and  the one given in Ref.\cite{Buzzegoli:2017cqy} for  $C_{\Omega}$.  The coefficient $C_{\Omega}$ denotes that the pseudoscalar condensation can be induced from the vorticity and acceleration by the inner product $\varepsilon\cdot \omega$. This is very relevant to the heavy-ion collision in which  the huge vorticity and acceleration are present. Pseudoscalar condensation can influence the neutral pion's production rate or bulk flow, we will postpone this interesting  study in the future.

{\it Summary} ---
We have shown that the GCKT can be achieved in unifying the quantum kinetic theory for  both massless and massive Dirac fermions.
In the GCKT, we  choose four independent phase-space distribution functions --- chiral (righthanded and lefthanded)
distributions and  transverse magnetic moment distribution --- as  the primary variables. These four independent distributions
fulfill four coupled kinetic equations. At chiral limit, these equations totally decouple with each other and the GCKT is reduced
into the CKT in a trivial way.

We apply this formalism to discuss the spin polarization and charge separation and find the dynamical magnetic-moment distribution will induce
 extra contribution compared to the massless fermions.  We also determine the specific solution in global equilibrium under vorticity and electromagnetic field.
 We find that  the electric or magnetic moment distributions can be induced by both vorticity or electromagnetic field up to  the first order. We also find the  pseudoscalar condensation can be generated by both vorticity or electromagnetic field up to  the second  order.

\textit{Acknowledgments.} ---
This work was supported in part by National Natural Science Foundation of China  under Nos. 11890710, 11890713, 12175123
and the Major Program of  Natural Science Foundation of Shandong Province under No. ZR2020ZD30.

\end{document}